# *-Quantum Effects or Theoretical Artifacts? A Computational Reanalysis of Hydrogen's High-Pressure Phase Stability and Properties*

Stefano Racioppi*,[1,2] Eva Zurek*[2]

[1]Department of Materials Science and Metallurgy, University of Cambridge, Cambridge CB30FS, United Kingdom.
[2]Department of Chemistry, State University of New York at Buffalo, Buffalo, New York 14260-3000, USA.

**ABSTRACT**. Experimental and theoretical studies of the high-pressure phases of hydrogen are highly sensitive to methodological choices. We compare the cold phase diagram of hydrogen calculated between 400-700 GPa with meta-GGA functionals ($R^2$SCAN and SCAN0) with the commonly employed GGA-PBE functional. Molecular phases are stabilized over the atomic phase to higher pressures with the meta-GGA, in closer agreement with diffusion Monte Carlo calculations. $R^2$SCAN phonon spectra show that the dynamical instabilities and anharmonic signatures predicted at the GGA level vanish, indicating that such effects may partly arise from functional deficiencies rather than quantum nuclear effects. Bonding analysis reveals that PBE artificially weakens intramolecular H–H bonds and enhances intermolecular interactions through charge delocalization, whereas meta-GGA preserves a more localized molecular character.

Ninety years ago, Bernal suggested that hydrogen, the lightest and most abundant element in the Universe, could be metallized under pressure. Wigner and Huntington predicted metallization in the solid at 25 GPa [1], though dynamic compression at 140 GPa and ~3000 K [2]. Though toroidal diamond anvil cells [3], hydrogen metallization in a crystalline sample at low temperature has not been unambiguously confirmed. Electrical conductivity and Raman measurements in static compression experiments have provided evidence for semi-metallic behavior at 350 GPa, with a possible transition to a metallic state above 440 GPa and 100 K, [4] while infrared optical absorption has hinted that the molecular solid metallizes near 425 GPa and 80 K [5]. Further measurements and improved theoretical models are required to conclusively establish the metallization of hydrogen at high pressure (HP) and low temperature conditions.

The experimental characterization of hydrogen's HP atomistic structure is extremely challenging, as the lightest element possesses an exceptionally low X-ray scattering cross section and a high Debye-Waller factor. A recent study employing nano-focused synchrotron X-ray probes suggested a post-*hcp* model between 212-245 GPa at room temperature. This structure consists of graphene-like layers where each hexagon is made of three $H_2$ dimers separated by layers of spherically disordered $H_2$ molecules, [6] resembling features previously proposed for phase IV of hydrogen [7–9]. Notably, the formation of graphene-like layered motifs in dense hydrogen has long been hypothesized [10,11].

Theory likewise continues to be challenged by the lightest of all atoms. Within Density Functional Theory (DFT), the workhorse of numerical quantum chemistry and band structure methods, computations involving hydrogen are plagued by the artificial interaction of the single electron with itself, due to the reliance on non-exact functionals of the electron density [13]. The main limitations include delocalization and static correlation errors, [12] together with the incorrect asymptotic behavior of the exchange-correlation potentials [13], which are amplified in the case of hydrogen. Even in an isolated H atom the binding (ionization) energy calculated by one of the most commonly employed Generalized Gradient Approximation (GGA) functionals, PBE, [14] can be off by as much as 50% compared to the experimental value (13.6 eV). This error stems from the estimation of repulsive electron-electron interactions within GGA, which are present despite the atom being mono-electronic [15]. For many-electron systems, like $H_2$ molecules, only high-level quantum mechanical calculations, [16–18] including Quantum Monte Carlo (QMC) [19] and coupled-cluster, [20] can achieve a truly accurate description. However, these exact methods incur a computational cost that scales exponentially with the system size, making them virtually inaccessible for studies of most real systems. Nonetheless, QMC methods were successfully used to train machine-learning potentials, which notably speeded up the exploration of dense hydrogen's phase diagram [21–23].

A practical strategy to improve the accuracy of computations on many-electron systems, including potentially costly geometry optimizations of extended structures, is the development and benchmarking of semi-local DFT functionals, which go beyond the GGA approximation, but keep the computational expense tractable [24,25]. Meta-GGA functionals [26] bring a substantial improvement by employing the kinetic energy density to provide a better description of the correlation energy, and by approaching the correct gradient expansions in the exchange and correlation energies for slowly varying densities (*i.e.*, metals). [27,28] As a consequence, meta-GGAs yield better band-gaps (severely underestimated by GGAs) and bonding properties (bond lengths, interaction strengths, and dissociation curves); [29,30] and in many cases, at almost GGA computational cost [31]. In a parallel vein, functionals accounting for van-der-Waals interactions have been shown to improve calculations for warm dense hydrogen. [32–34]

Herein, we perform state-of-the-art calculations of the cold (0 K) HP phase diagram of hydrogen between 400-700 GPa optimized at three levels of theory – PBE, R$^2$SCAN and SCAN0 – to highlight the impact of the exchange-correlation functional on the phases' relative enthalpies, without zero-point energy contributions, at 0 K (Figure 1). We focused on structures that have been studied as likely candidates including the molecular *Cmca*-12 [35], *Cmca*-4 [36] and *C*2/*c* (phase-III) [35] phases (whose dynamic stability was thought to be governed by anharmonic effects), [37,38] and on the atomic $I4_1/amd$ [35] phase, which assumes the Cs-IV structure type. [39] In addition, we searched for previously unreported phases of pure hydrogen at 500 GPa using the evolutionary algorithm XtalOpt [40,41], but none were found.

The GGA functional (PBE) finds the enthalpy of the atomic $I4_1/amd$ phase to fall below the enthalpy of *Cmca*-4 at ~507 GPa, comparable to results obtained previously with the same functional (490 GPa) but using a different code [42]. In contrast, previous diffusion quantum Monte Carlo calculations (DMC), [43] predicted that $I4_1/amd$ becomes the ground state at 374 GPa. Because of the fixed-node approximation [44] used by DMC, these calculations relied on the local density approximation (LDA) of the orbital energies and initial geometry, which inevitably affects the final result of the Monte Carlo simulation [45]. All calculated phases were predicted to be metallic within PBE.

Moving to the meta-GGA level, R$^2$SCAN predicts the *Cmca*-4 phase as being more stable than *Cmca*-12 above 428 GPa and up to 675 GPa, before transforming into the atomic $I4_1/amd$ phase (Figure 1). At this level of theory, the ground state phases were predicted to be metallic. However, the *C*2/*c* phase, which is just a couple of meV/atom less stable than the ground state (Figure S1), possessed a band gap of 0.41 eV at 400 GPa, which closes below 500 GPa. A recent series of DMC calculations, this time based on PBE geometries and orbital energies, instead of LDA [46], similarly predicted a transition from the *Cmca*-12 phase in this pressure regime (447 GPa), but now to the atomic $I4_1/amd$ phase, and not to *Cmca*-4. Is it possible to go beyond meta-GGA? In principle, hybrid and higher level functionals would be ideal for studying hydrogen computationally, since the inclusion of Hartree-Fock exact exchange has a significant impact on the electronic energies and optimized geometries [12]. Unfortunately, using such high levels of theory in periodic systems, including HP hydrogen, is hindered by the huge computational cost required to calculate the exact exchange with plane-wave based codes. We therefore performed single point calculations of the electronic energies of the *C*2/*c*, $I4_1/amd$, *Cmca*-4 and *Cmca*-12 phases from 400-700 GPa using the meta-hybrid-GGA functional SCAN0 on the R$^2$SCAN optimized geometries (Figure 1) [47].

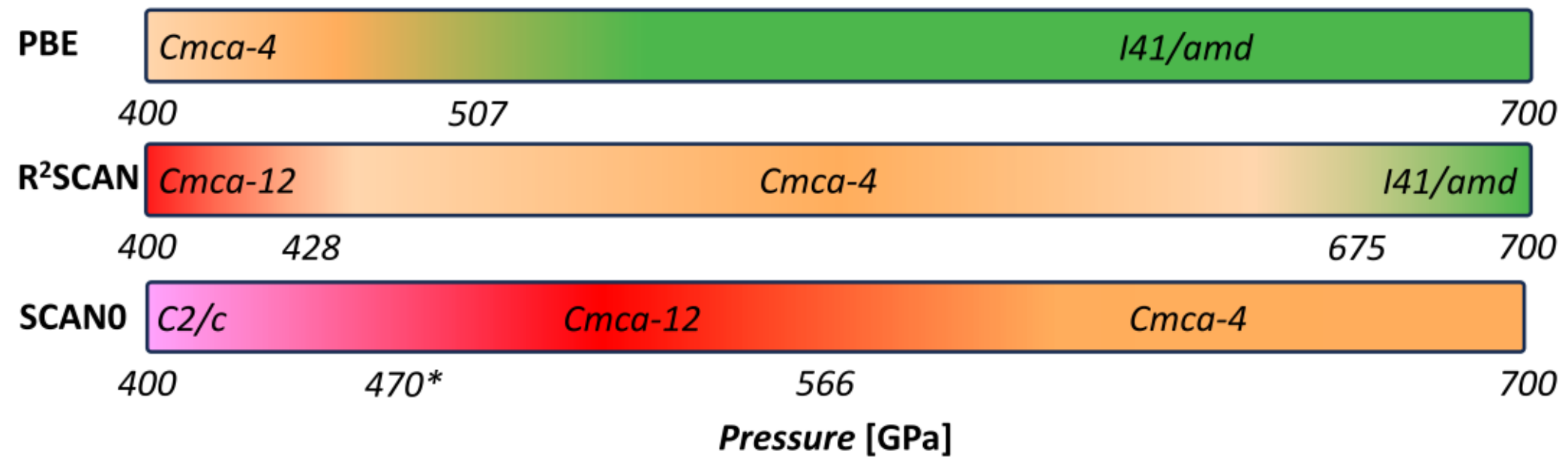

**Figure 1.** Cold (0 K) phase diagrams of HP hydrogen, using the functionals: (1) PBE (GGA); (2) R$^2$SCAN (meta-GGA); and (3) SCAN0 (meta-hybrid-GGA). The SCAN0 enthalpies were calculated on the R$^2$SCAN geometries. At 470 GPa (*), SCAN0 predicts hydrogen's metallization, in agreement with experimental observations [4,5].

All forces estimated by SCAN0 were below $10^{-13}$eV/Å, suggesting that the structures would not change much upon re-optimization. However, the SCAN0 internal energies differed from those computed with R$^2$SCAN. The most striking result is that at 400 GPa the *C*2/*c* phase is now the ground state with a band gap of 1.75 eV, which closes at ~470 GPa upon transition to *Cmca*-12 (Table S1). This insulator-to-metal transition is now in good agreement with the infrared data obtained at 80 K by Loubeyre and coworkers (transition observed above 425 GPa), associated with a structural transition from *C*2/*c* to *Cmca*-12 [5]. Experiments concluded that hydrogen remains as *C*2/*c* up to ~450 GPa and 100 K, where the Raman signals disappear, suggesting the transition to a metallic molecular state [4]. Therefore, the hybrid meta-GGA level of theory gives the best agreement with state of the art experimental data. At higher pressures *Cmca*-12 transforms into *Cmca*-4 at ~566 GPa. By extrapolation of the meta-hybrid GGA enthalpies, the stabilization of the atomic $I4_1/amd$ phase is estimated to occur above 850 GPa, beyond any cold experiment performed thus far on hydrogen.

Within this first section, we aimed to highlight the sensitivity of the predicted phase diagram of hydrogen to the DFT level of theory. These results suggest that similar sensitivity would be observed for other quantum states of matter where hydrogen interactions are important, such as superionicity, [48] or the formation of superconducting hydrogen cages. [49] Moreover, we observed that increasing the level of theory in static (0 K) calculations pushes the range of thermodynamic stability of the atomic $I4_1/amd$ phase progressively towards higher pressures, favoring the molecular phases.

In the second part of this work, we aim to quantify how different DFT approximations impact the lattice dynamics of compressed hydrogen. Previous calculations at the BLYP level of theory, [38,50] where anharmonic effects were included to relax the geometries and correct the enthalpies, predicted *Cmca*-12 as the ground state at 410 GPa, followed by a transformation to the atomic phase at 577 GPa, [35,38] without passing through the *Cmca*-4 phase. Similarly, other studies using the PBE functional predicted that *Cmca*-4 becomes dynamically unstable at 450 GPa. [37,51] However, the unstable phonons could be renormalized by anharmonic effects calculated with the stochastic self-consistent harmonic approximation (SSCHA) [37,51]. Our calculations give a different picture. Specifically, the meta-GGA functional R$^2$SCAN does not yield imaginary phonon modes for *Cmca*-4, even at 700 GPa (Figure 2). Below we shed light on the origin of these discrepancies, by investigating the functional-dependence of dynamic stability.

Figures 2a-d compare the PBE and the R$^2$SCAN harmonic phonon spectra for the four phases at 700 GPa. For the atomic phase the phonon bands calculated with the two functionals are similar, apart from a slight hardening of the modes below 1500 cm$^{-1}$ along $\Gamma \rightarrow$ X and at N with the meta-GGA. Though the $I4_1/amd$ phase has been predicted to be highly harmonic [38,50,52], the inclusion of anharmonic quantum nuclear effects (QNE) at 0 K, within PBE, produced softer phonon modes [53]. Those calculations were, however, performed near the PBE phase transition's pressure. The meta-GGA functional calculates a slightly larger volume per atom in $I4_1/amd$ compared to GGA, especially at lower pressures (Figure S2), but the effect on the phonons is not as evident as for the molecular phases.

We are not the first to notice that the PES of $I4_1/amd$ can be adequately calculated even by lower levels of theory. Ref [38] compares the calculated PES of $I4_1/amd$ as a function of the *c/a* ratio between LDA, GGA and QMC. Surprisingly, for small atomic volumes (*i.e.*, high-pressures) LDA and GGA predict energy profiles that almost match that of QMC. This is likely due to the transition from covalent molecule (localized electron pair) to atomic metal (free electron gas-like), which can be fairly well described also by the LDA.

The geometries and the phonon spectra of the *Cmca* and the *C*2/*c* molecular phases are the most affected by the choice of functional. Within the harmonic approximation, PBE (Figure 2b) and BLYP [37,38] find *Cmca*-4 to be dynamically unstable at 700 GPa. At the BLYP level of theory, the instability arises already at ~400 GPa [37], while R$^2$SCAN predicts neither dynamic instabilities nor soft phonon modes that could be signatures of anharmonicity (Figure 2b). As we will discuss later, this difference boils down to the ability of meta-GGA to correctly calculate intra and intermolecular hydrogen interactions [54]. A similar hardening of the phonons is evident both in *Cmca*-12 (dynamically stable within PBE) and in *C*2/*c* (dynamically unstable within PBE) once the meta-GGA functional is used (Figure 2c-d).

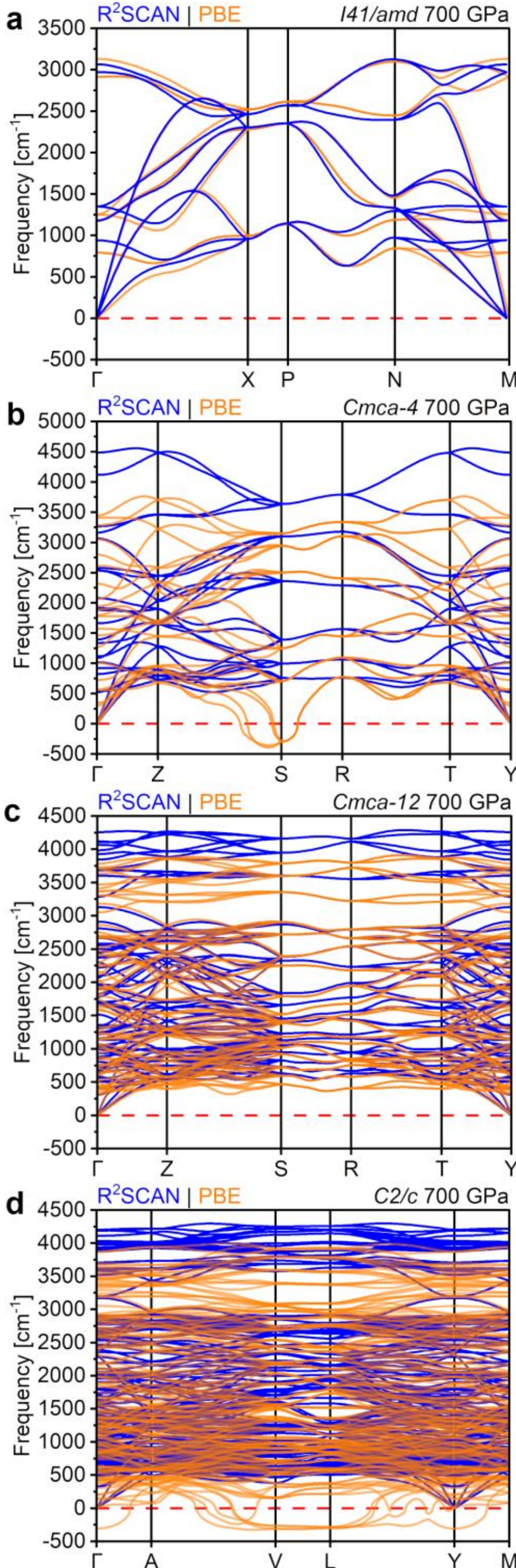

**Figure 2.** Phonon band structures of the optimized atomic $I4_1/amd$, and molecular *Cmca*-4, *Cmca*-12 and *C*2/*c* phases of hydrogen, calculated at 700 GPa using both the PBE (orange lines) and the R$^2$SCAN (blue lines) functionals.

Interestingly, R$^2$SCAN calculates a volume expansion for all the phases, resembling the expansion caused by QNE (Figure S2) [37]. With the SSCHA this expansion results from the QNE-induced elongation of the H-H bonds from 0.793 Å to 0.837 Å at 414 GPa [37]. In contrast, within R$^2$SCAN the volume expansion is due to an increase in the distance between the molecules, due to a higher Pauli repulsion, while the H-H distances shorten and strengthen (0.759 Å at 400 GPa). The crucial effect of explicitly introducing the dependence on $\nabla^2\rho$ in the functional becomes even more evident in quantum molecular dynamics calculations (Figure S3). *NPT* simulations on the *Cmca*-4 phase at 700 GPa and 300 K with the PBE functional melts the solid, as shown by the $g(r)$ profile, or radial distribution function (Figure S3), in agreement with the imaginary modes obtained by the harmonic phonons (Figure 2b). On the other hand, the $g(r)$ obtained with the meta-GGA functional is typical of a crystalline system, even at room temperature (Figure S3).

To quantify the differences between functionals, let's focus on the intramolecular interactions in molecular hydrogen (Figure 3). In the relaxed *Cmca*-4 phase at 700 GPa, used here as a general example, the intramolecular H-H distance equals 0.836 Å and 0.752 Å for PBE and R$^2$SCAN, respectively. Extracting a rigid molecular unit of $H_2$ from the unit cell (and placing it in a 10 Å$^3$ box, without performing further relaxations) and calculating the difference between its lowest unoccupied molecular orbital (LUMO) and its highest occupied molecular orbital (HOMO), PBE and R$^2$SCAN predict gaps of 9.73 eV and 10.98 eV, respectively. Scanning the energy profile of the extracted $H_2$ molecule as a function of the H-H distance shows that the meta-GGA functional finds the minimum of the isolated molecule very close to that in *Cmca*-4 at 700 GPa (Figure 3). In other words, for R$^2$SCAN, the *Cmca*-4 phase is a pure, almost unperturbed, molecular crystal. In fact, by re-optimizing the geometry of the isolated molecule, the H-H bond shortens by only 0.01 Å, and the band gap (HOMO – LUMO difference) decreases by less than 0.1 eV (Figure 3).

In contrast, the energy profile calculated with PBE reveals that the H-H distance in the extracted isolated molecule differs substantially from that in the crystal, suggesting that the molecular nature of $H_2$ is perturbed within the *Cmca*-4 phase. PBE weakens the H-H bond when the molecule is embedded in the crystal lattice, resulting in an elongation of almost 0.1 Å. GGA functionals tend to over delocalize the electron density, producing more metallic periodic systems, and causing the famous underestimation of band (and HOMO – LUMO) gaps. PBE relaxation of the isolated molecule contracts the H-H distance to 0.750 Å, and the HOMO – LUMO increases to 10.15 eV, in-line with the $R^2$SCAN calculation.

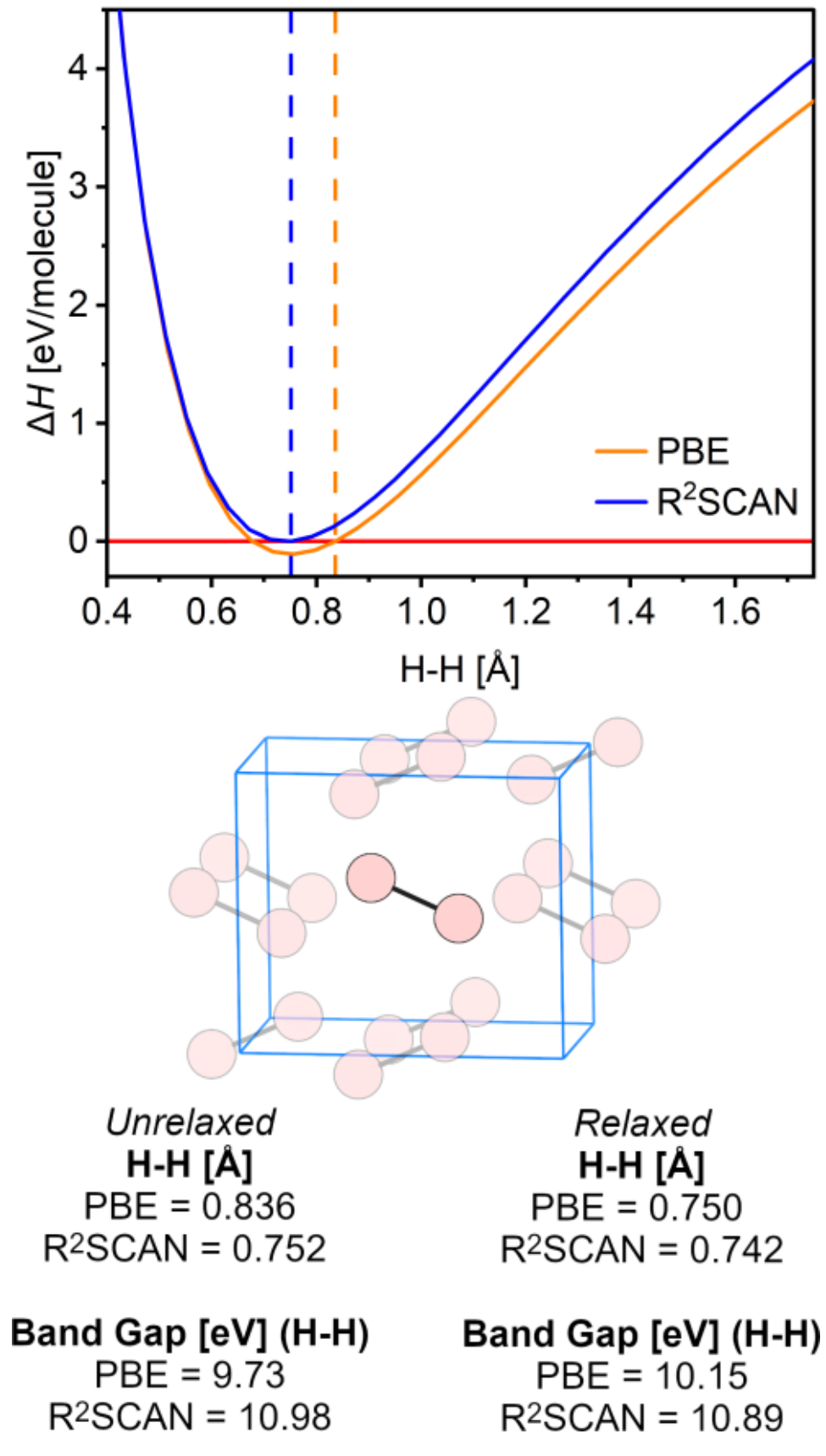

**Figure 3**. (Top) Energy profiles of isolated $H_2$ as a function of the intermolecular distance, calculated with $R^2$SCAN and PBE. Dashed lines correspond to the intermolecular distances in *Cmca-4* at 700 GPa. The zero is defined by the energy of the unrelaxed $H_2$ molecule extracted from the crystal. (Bottom) The interatomic distances and HOMO – LUMO gaps of the unrelaxed (extracted from *Cmca*-4 at 700 GPa) and the relaxed hydrogen molecule.

The effects of electron density delocalization calculated by PBE in *Cmca*-4 can be further quantified by means of the Quantum Theory of Atoms in Molecules (Section S3) [40]. In this context, the value of the electron density $\rho(\mathbf{r})$ at the intramolecular bond critical point, bcp, becomes a useful descriptor to benchmark the effect of the functional on the bond strength, while the value of the Laplacian, $\nabla^2\rho(\mathbf{r})$, quantifies the degree of localization of the electron density at the bonding critical point. The values of $\rho(\mathbf{r})$ and $\nabla^2\rho(\mathbf{r})$ calculated with PBE are quite low, signifying a more diffuse electron density and flatter curvature compared to $R^2$SCAN, and demonstrating that the GGA produces weaker H-H bonds. However, PBE counterbalances the weakening of the intramolecular interaction by delocalizing the charge density among the adjacent molecules and strengthening the intermolecular interactions (Figure S4). This phenomenon also affects the density of states at the Fermi level (Figure S5), which in the case of PBE, is almost 30% higher than in $R^2$SCAN. This marked difference in the electronic structure and electron density calculated by the GGA and meta-GGA functionals is not observed in the atomic *I*4*₁/amd* phase, which instead, is well described by both functionals as a metallic atomic phase (Figure S4-S5).

To conclude the bonding analysis, we used the Periodic Energy Decomposition Analysis (Section S4) [40,55] to find the electrostatic (ionic), orbital (covalent) and Pauli repulsion contributions (Figure S6) towards the intermolecular interaction between a hydrogen molecule and its surroundings in *Cmca*-4. Three different scenarios were considered: (1) from the PBE energy calculated in the PBE relaxed structure; (2) from the $R^2$SCAN energy calculated in the PBE relaxed structure; and (3) from the $R^2$SCAN energy calculated in the $R^2$SCAN relaxed structure. Since the unit cell and the atomic positions are the same, the energy difference between (1) and (2) reveals the difference between the two functionals. $R^2$SCAN calculates weaker interactions between the reference $H_2$ molecule and the crystalline environment. The higher Pauli repulsion in (2) and lower orbital interaction with the adjacent molecules, are the terms responsible for this effect, since the charge is less delocalized and more concentrated in the $H_2$ intramolecular bond with the meta-GGA functional (Figure S6). In fact, in (3), where the unit cell and atomic coordinates are optimized using the meta-GGA functional, the $H_2$ bond shortens, while the unit cell volume expands. In energy terms, this is translated into a drop of the Pauli repulsion between molecules, but also a decrease of the intermolecular strength. Therefore, the volume expansion in *Cmca-4* calculated by $R^2$SCAN is the way that the system adapts to lowering the intermolecular Pauli repulsion.

In conclusion, we analyzed the effect of GGA, meta-GGA, and meta-hybrid-GGA functionals on the phase stability, lattice dynamics, and bonding of hydrogen between 400 and 700 GPa. Increasing the level of theory progressively stabilizes the molecular phases relative to the atomic *I*4*₁/amd* phase, yielding a metallization pressure that is in close agreement with current experimental observations [5]. We further show that the phonon instabilities and apparent anharmonic signatures predicted for molecular phases at the GGA level disappear within $R^2$SCAN, indicating that some of these effects may originate from deficiencies in the exchange-correlation functional rather than from genuine quantum nuclear motion. Although anharmonic effects remain important in dense hydrogen, our results demonstrate that an accurate description of the underlying potential energy surface is a prerequisite for reliably assessing its phase stability, lattice dynamics, and bonding strength.

## ACKNOWLEDGMENTS

Funding for this research was provided by the *Center for Matter at Atomic Pressures* (CMAP), a National Science Foundation (NSF) Physics Frontier Center, under Award PHY-2020249. Partial funding for this research is provided by the U.S. Department of Energy, Office of Science, Fusion Energy Sciences funding the award entitled *High Energy Density Quantum Matter*, under Award No. DE-SC0020340. Calculations were performed at the Center for Computational Research at SUNY Buffalo (http://hdl.handle.net/10477/79221).

**Supplementary Material**

## S1. Computational Details

### Crystal Structure Prediction

The open-source evolutionary algorithm XtalOpt [1,2] version 13.0 was employed for crystal structure prediction (CSP). Two CSP searches were performed at 500 GPa. The initial generation consisted of 300 random symmetric structures that were created by the RandSpg algorithm. [3] The number of H atoms in the unit cell considered in each run was equal to 1, 2, 3, 4, 6, 8, 12, 16, 20, 24 and 32. Duplicate structures were identified and removed from the breeding pool using the XtalComp algorithm. [4] The total number of generated structures was equal to 1000 per run. Each structure search followed a multi-step strategy, with three subsequent optimizations with increased level of accuracy, followed by a final accurate single point calculation. For the CSP searches, calculations were performed using Density Functional Theory (DFT) with the Vienna Ab Initio Simulation Package (VASP), version 6.4.2. [5] The meta-GGA $R^2$SCAN-L [6] exchange-correlation functional was employed. The projector augmented wave (PAW) method [7] was used to treat the core states in combination with a plane-wave basis set with an energy cutoff of 600 eV for the geometry optimizations. The non-spherical contributions related to the gradient of the density in the PAW spheres were included. The *k*-point meshes were generated using the Γ-centered Monkhorst−Pack scheme, [8] and the number of divisions along each reciprocal lattice vector was selected so that the product of this number with the real lattice constant was greater than or equal to 60 Å. The accuracy of the energy convergence was set to increase from $10^{-5}$ eV, for which the norms of all the forces calculated during the relaxations were smaller than $10^{-3}$ eV/Å. A Gaussian smearing was used at each optimization step with a sigma of 0.02 eV.

### Electronic Structure and Topology

We performed periodic DFT calculations using VASP, version 6.4.2. [5] The $R^2$SCAN [9] and PBE [10] exchange-correlation functionals were employed for the geometry optimizations and calculations of the electronic structure of the known structures of hydrogen. Geometry optimizations of the candidate structures were performed between 400 GPa and 700 GPa, using 100 GPa increments. The projector augmented wave (PAW) method was used, [7] with a cutoff of 700 eV. The *k*-point meshes were generated using the Γ-centered Monkhorst−Pack scheme, [8] and the number of divisions along each reciprocal lattice vector was selected so that the product of this number with the real lattice constant was greater than or equal to 60 Å. For the geometry optimizations, the accuracy of the energy convergence was set to $10^{-6}$ eV and the norms of all the forces were smaller than $10^{-3}$ (eV/Å). The Methfessel-Paxton smearing method was adopted in the geometry optimizations, [11] while the tetrahedron smearing method was used to calculate the electronic properties of the optimized geometries. [12] The non-spherical contributions related to the gradient of the density in the PAW one-center terms were included. For the projected density of states, the Wigner-Seitz radius of H was recalculated for each phase and at each pressure point form the volume of the unit cells. The topological analysis of the electron density, based on the Quantum Theory of Atoms in Molecules (QTAIM), [13] was performed using the Critic2 code. [14] Phonons in the harmonic approximation were determined with the Phonopy package [15] using supercells of 3x2x3 for *Cmca-12*, 4x3x3 for *Cmca-4* and 4x4x2 for $I4_1/amd$, based on the conventional unit cells of the optimized structures. A finite displacement of 0.003 Å was used, and the accuracy of the energy convergence was set to $10^{-8}$ eV. Single point energies were also calculated at the meta-hybrid GGA level of theory with the SCAN0 [16] functional on the $R^2$SCAN optimized geometries maintaining the same computational parameters of accuracy, cutoff, smearing (tetrahedron) and *k*-point mesh.

### Molecular Dynamics

Molecular dynamics simulations were performed with VASP on the *Cmca-4* phase using an isothermal-isobaric (NPT) ensemble and the Langevin thermostat [17,18] at $T$ = 300 K and at $P$ = 700 GPa. Both the PBE and the $R^2$SCAN functionals were tested on 4x3x3 supercells (288 atoms). The duration of the simulation was set to 5 ps with 1 fs time steps. The projected augmented wave (PAW) method, [7] with a cutoff of 700 eV was used and the accuracy of the energy convergence was set to $10^{-5}$ eV. A Fermi smearing was employed.

**Energy Decomposition Analysis**

To evaluate the energy interaction terms within periodic energy decomposition analysis (PEDA), [19] the BAND package, version 2022.101, [20] was used on the *Cmca-4* hydrogen phase optimized with the VASP using the PBE and the $R^2$SCAN functionals at 700 GPa. Scalar relativistic effects were included using the Zero Order Regular Approximation (ZORA), [21] along with a double-ζ polarized (DZP) basis set. [22] The integration quality was kept equal to 'Very Good' for all calculations.

**S2. Thermodynamics and Molecular Dynamics**

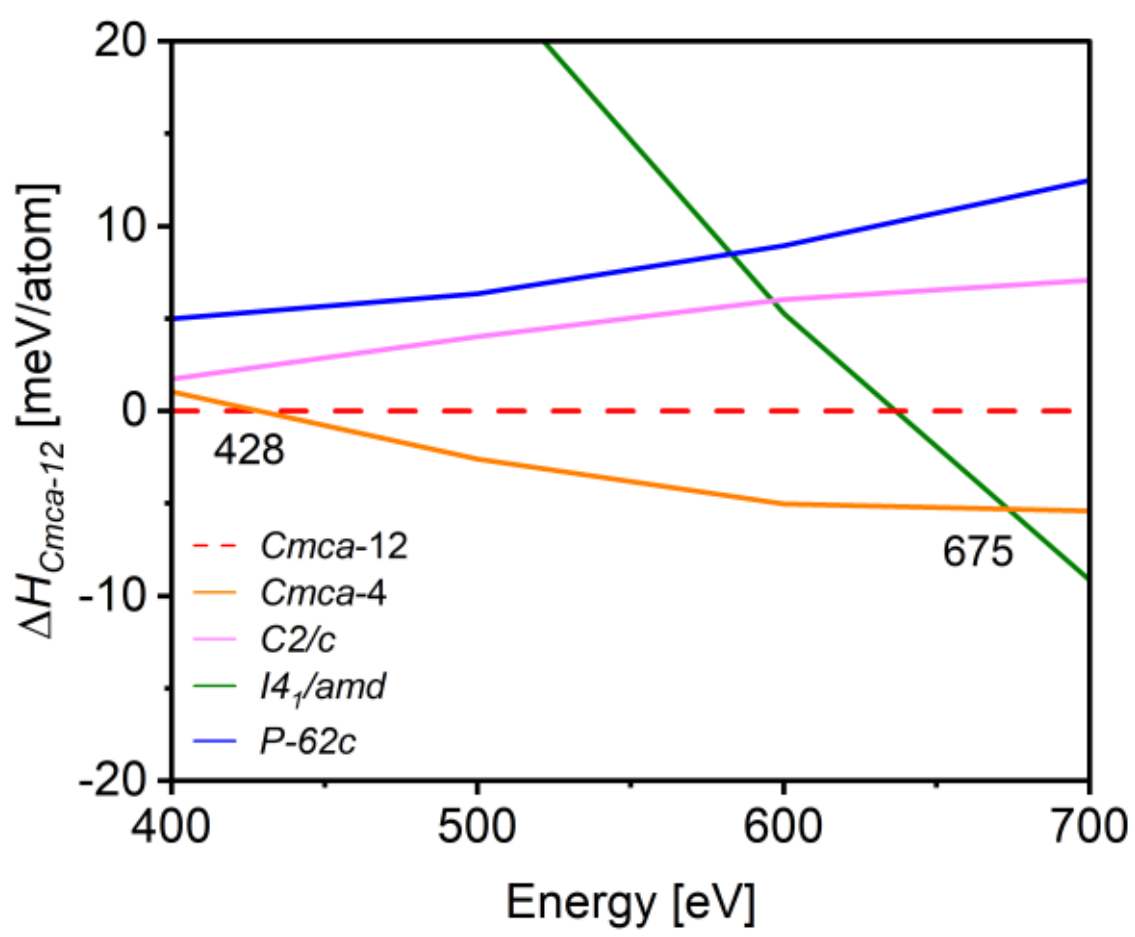

**Figure S1**. Enthalpy differences ($\Delta H = \Delta E + P\Delta V$) as a function of pressure for the high-pressure phases of hydrogen calculated with VASP using the meta-GGA exchange-correlation functional $R^2$SCAN, between 400 GPa and 700 GPa. The transition pressures are marked at 428 and 675 GPa. Notice that the previously reported *$P6_2/c$* structure, [23] refers in reality to the *P-62c* space group, since *$P6_2/c$* does not exist as a space group. [24]

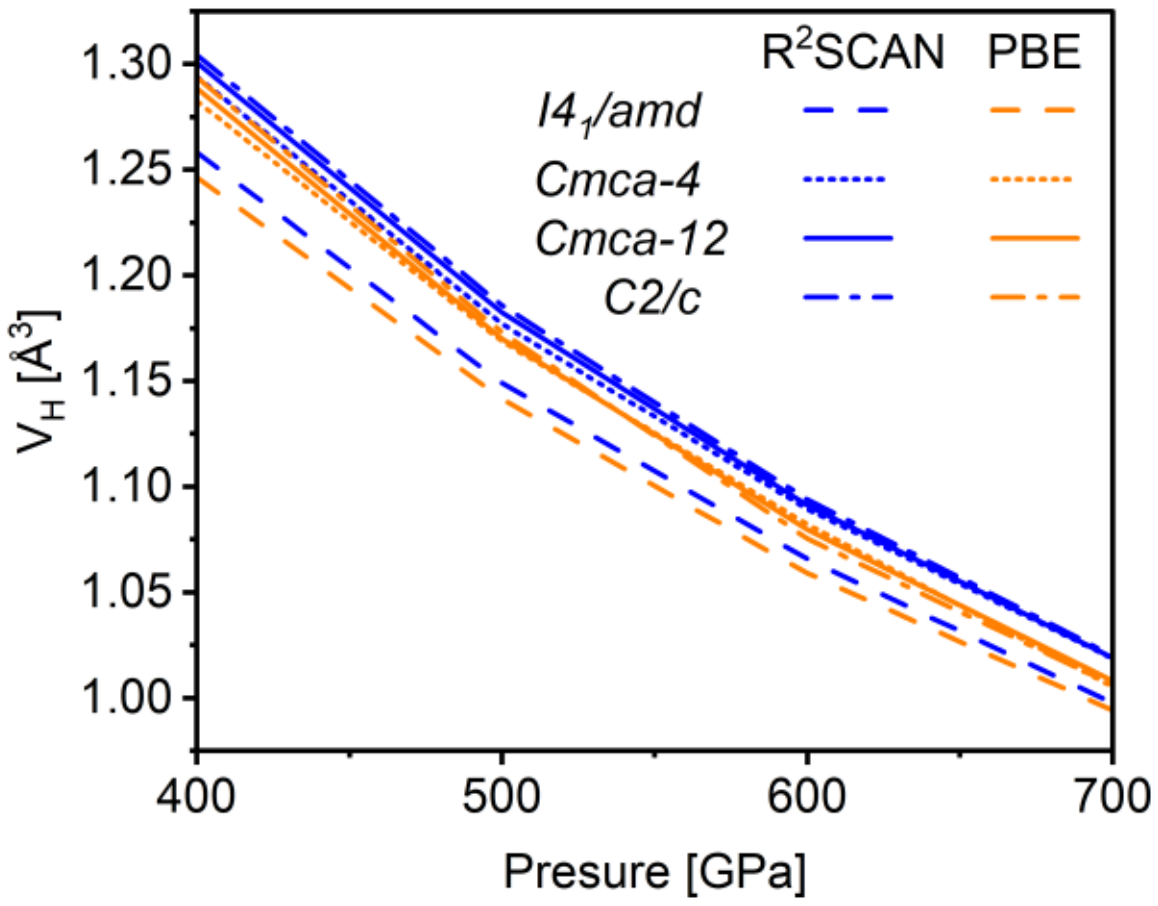

**Figure S2**. Atomic volumes per hydrogen atom as a function of pressure, calculated with VASP using the PBE and the $R^2$SCAN functionals.

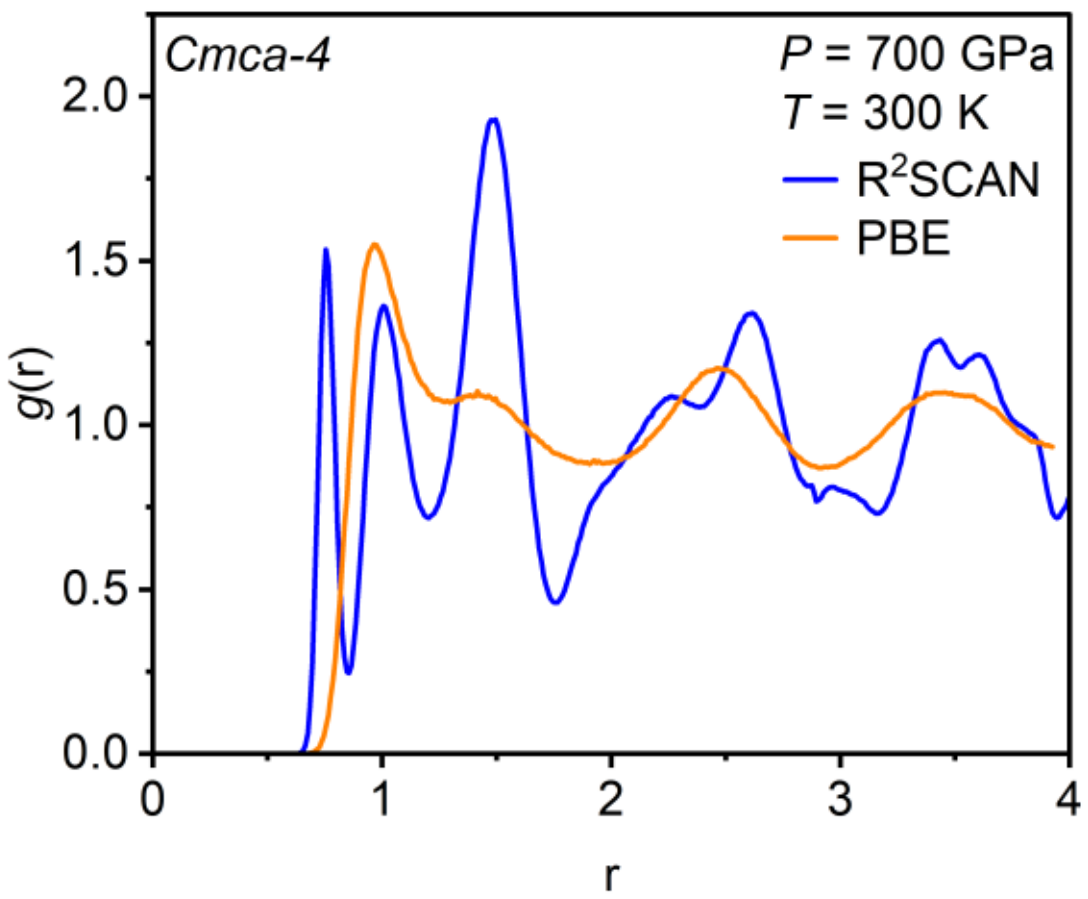

**Figure S3**. Pair distribution function, *g(r),* calculated via *NPT* molecular dynamics simulation on *Cmca*-4 at *T*=300 K and *P* = 700 GPa with $R^2$SCAN (blue line) and PBE (orange line). The *g(r)* profile calculated the with meta-GGA functional is that of a crystalline phase, while the one calculated with the GGA functional is that of a liquid phase.

**S3. Overview on the Quantum Theory of Atoms in Molecules (QTAIM).**

The QTAIM is the cornerstone method to study chemical bonding in real space, both from theoretical and experimental data. [25,26] QTAIM studies the local topological features of the electron density, $\rho(\mathbf{r})$, mapping critical points and evaluating chemical bonding and properties based on $\rho(\mathbf{r})$.

The bond critical points (bcp), in particular, are special points in the electron density where $\nabla\rho(\mathbf{r}) = 0$; and they are associated with the presence of bonding interactions between atoms. [26] The amount of electron density at the point $\rho(\mathbf{r})$ gives an estimate of the strength of the interaction between two atoms. [27] The Laplacian of the electron density, $\nabla^2\rho(\mathbf{r})$, quantifies the degree of localization and curvature of the electron density at the bcp. Generally, large and negative values of $\nabla^2\rho(\mathbf{r})$ correspond to covalent interactions, while $\nabla^2\rho(\mathbf{r}) > 0$ is found in hydrogen-bonds or van der Waals interactions. The bond ellipticity, $\varepsilon$, is related to the deviation of charge distribution from cylindrical symmetry ($\sigma$-type of interaction) at a bcp. Values of $\varepsilon \sim 0$ correspond to interactions having cylindrical symmetry.

The total energy density at the bcp, named H, is the sum of G + V, the kinetic energy and the potential energy densities at the bcp. The more negative the H, the stronger the interaction. The kinetic energy density G is evaluated following the Abramov approximation. [28,29] The potential energy density at the bcp is then calculated using the local form of the virial theorem, $2G + V = 1/4\nabla^2\rho(\mathbf{r})$, which will consequently be negative. The ratio $|V|/G$, is useful to characterize the type of interaction at the bcp. [26,30] A ratio $|V|/G < 1$ corresponds to a closed-shell interaction region (ionic), while $|V|/G > 2$ corresponds to a shared-shell interaction region, such as covalent a bond. Values between $1 < |V|/G < 2$ fall in the intermediate regions. [26]

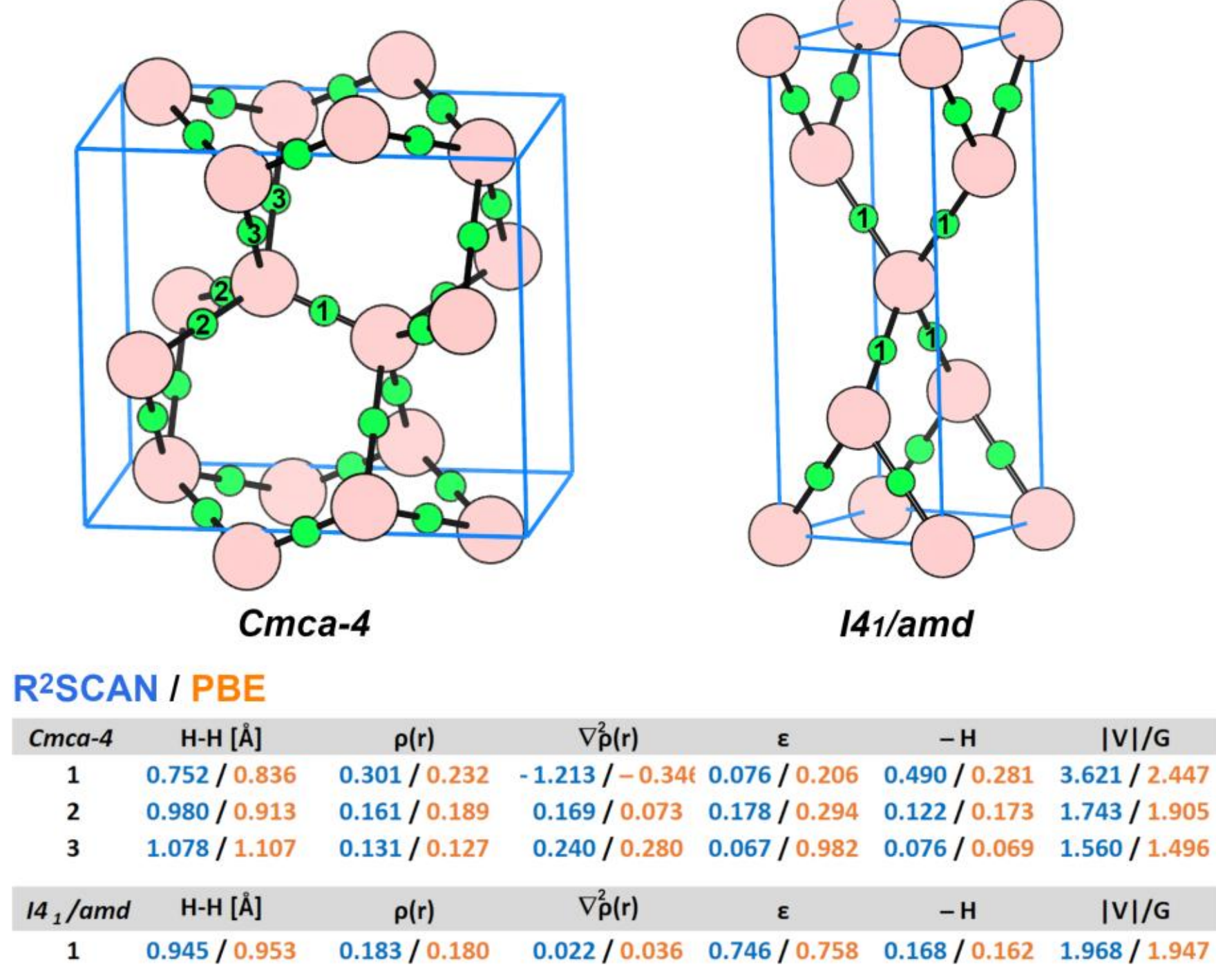

| Cmca-4 | H-H [Å] | ρ(r) | $\nabla^2\rho(r)$ | ε | – H | \|V\|/G |
|---|---|---|---|---|---|---|
| 1 | 0.752 / 0.836 | 0.301 / 0.232 | - 1.213 / – 0.34[illegible] | 0.076 / 0.206 | 0.490 / 0.281 | 3.621 / 2.447 |
| 2 | 0.980 / 0.913 | 0.161 / 0.189 | 0.169 / 0.073 | 0.178 / 0.294 | 0.122 / 0.173 | 1.743 / 1.905 |
| 3 | 1.078 / 1.107 | 0.131 / 0.127 | 0.240 / 0.280 | 0.067 / 0.982 | 0.076 / 0.069 | 1.560 / 1.496 |

| $I4_1/amd$ | H-H [Å] | ρ(r) | $\nabla^2\rho(r)$ | ε | – H | \|V\|/G |
|---|---|---|---|---|---|---|
| 1 | 0.945 / 0.953 | 0.183 / 0.180 | 0.022 / 0.036 | 0.746 / 0.758 | 0.168 / 0.162 | 1.968 / 1.947 |

**Figure S4**. (Top) Crystal structures of *Cmca*-4 and $I4_1/amd$ in their conventional unit cells, optimized with both the PBE and R$^2$SCAN functionals at 700 GPa. The green spheres mark the topological bond critical points, which are enumerated in accordance to the table below. (Bottom) Topological descriptors calculated using the QTAIM. H-H distance [Å]; ρ(**r**): electron density at the bcp [e/bohr$^3$]; $\nabla^2\rho(\mathbf{r})$: Laplacian of the electron density at the bcp [e/bohr$^5$]; ε: bond ellipticity; -H: total energy density at the bcp taken with the negative value [hartree/bohr$^3$]; $|V_b|/G_b$: potential/kinetic energy density ratio. [26]

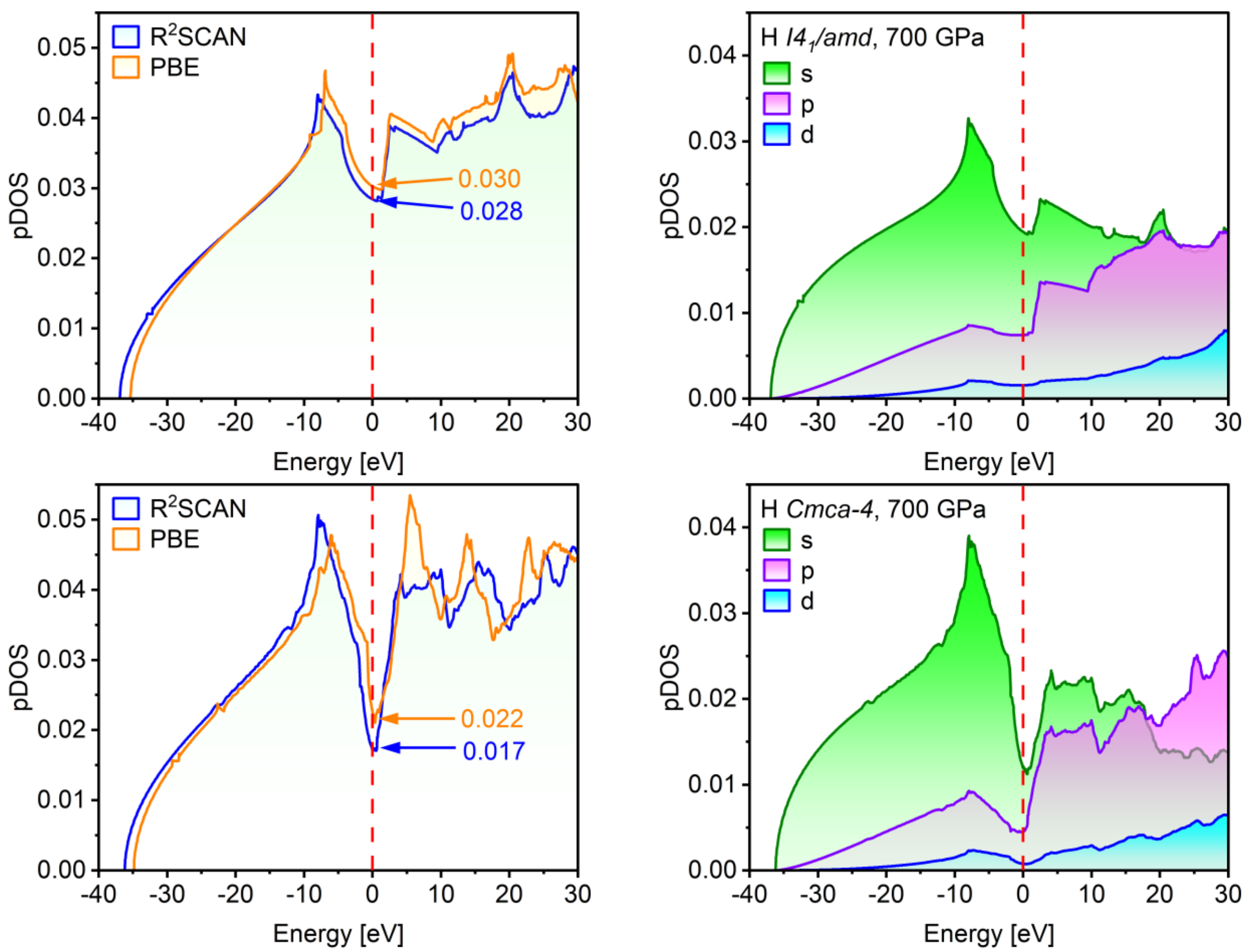

**Figure S5**. (Left) Total density of states (DOS) calculated for *I4₁/amd* (top) and *Cmca*-4 (bottom) at 700 GPa with VASP using the PBE and the $R^2$SCAN functionals. The values of the density of states per atom at the Fermi level ($g(E_F)$) are reported. The value of DOS at $g(E_F)$ calculated with PBE for the atomic phase is only 7% higher compared to the one calculated at the meta-GGA level of theory. On the other hand, for the molecular phase *Cmca*-4, the DOS at $g(E_F)$ calculated with PBE is almost 30% higher than with $R^2$SCAN. (Right) Projected density of states (pDOS) calculated for *I4₁/amd* (top) and *Cmca*-4 (bottom) at 700 GPa with VASP using the $R^2$SCAN functional.

**S4. Periodic Energy Decomposition Analysis (PEDA)**

To better understand the character of the intermolecular interactions between a molecule of $H_2$ in *Cmca*-4 with the rest of the crystal, we performed an energy decomposition analysis (EDA) using the Ziegler-Rauk variant, [31] which decomposes the total interaction energy ($\Delta E_{Int}$) between two prepared fragments into chemically meaningful terms. The term "prepared" means that the atoms in the moieties that are compared possess the same local coordinates as in the final product (Figure S6). Notice that all the interaction energies, $\Delta E_{Int}$, are positive, since they account only for the electronic energy associated with the reaction (they do not consider the *PV* contribution).

In the EDA, the total interaction energy, $\Delta E_{Int}$, is decomposed into a Pauli repulsion energy, $\Delta E_{Pauli}$, which is usually destabilizing; an electrostatic term, $\Delta E_{Elec}$, which can quantify the ionic character of an interaction, and finally an orbital term, $\Delta E_{Orb}$, which accounts for the covalent character of the total interaction. Often, the Pauli repulsion energy and the electrostatic energy are coupled together in a term named the steric interactions, $\Delta E_{Steric}$, which combines all these two components in a chemically intuitive term.

The total interaction energy, $\Delta E_{Int}$, calculated by EDA should not be confused with the $\Delta E$ term in the calculation of the enthalpy of formation, $\Delta H$. In the first case, the electronic energy of interaction is evaluated upon the reaction between two moieties possessing the same geometry as in the final product. In the second, $\Delta H$ is evaluated from the isolated reactants at their ground states, which and are comprised of the thermodynamic terms $\Delta E$ and $P\Delta V$.

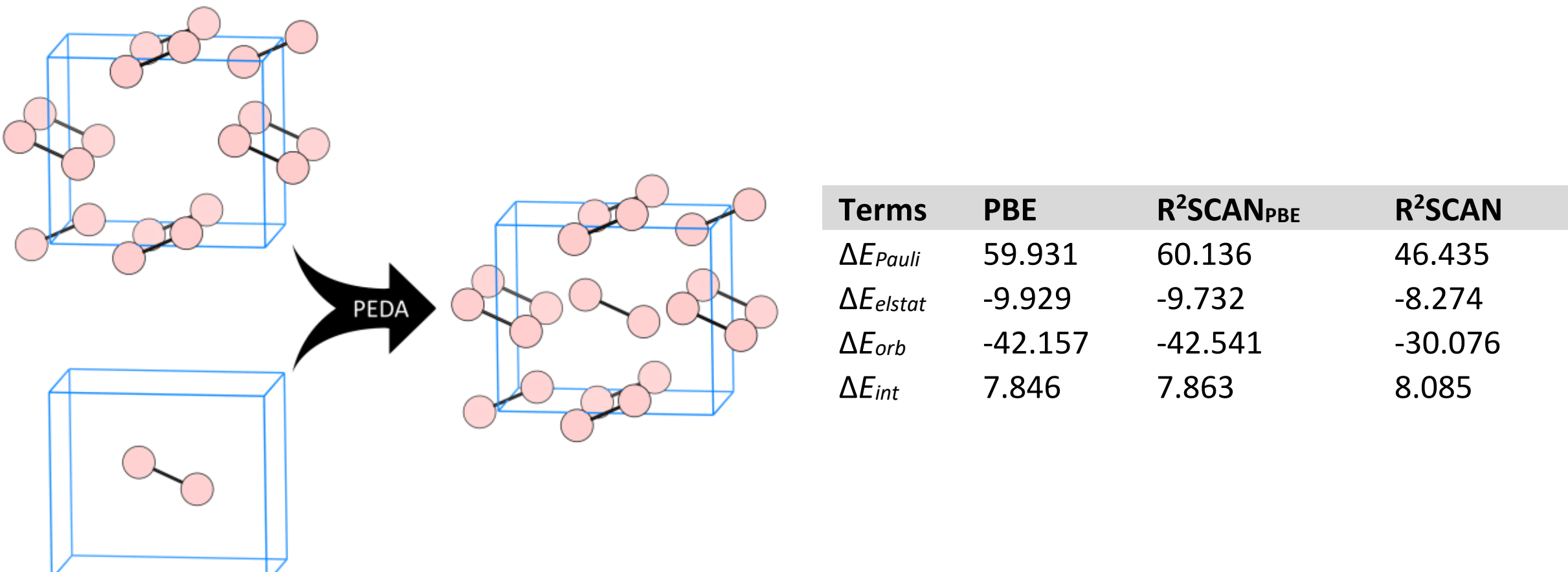

| Terms | PBE | $R^2SCAN_{PBE}$ | $R^2SCAN$ |
| --- | --- | --- | --- |
| $\Delta E_{Pauli}$ | 59.931 | 60.136 | 46.435 |
| $\Delta E_{elstat}$ | -9.929 | -9.732 | -8.274 |
| $\Delta E_{orb}$ | -42.157 | -42.541 | -30.076 |
| $\Delta E_{int}$ | 7.846 | 7.863 | 8.085 |

**Figure S6**. (**Left**) Scheme of the energy decomposition analysis. (**Right**) PEDA energy terms. PBE = PBE calculation on the PBE optimized structure at 700 GPa; $R^2SCAN_{PBE}$ = $R^2SCAN$ calculation on the PBE optimized structure at 700 GPa; $R^2SCAN$ = $R^2SCAN$ calculation on the $R^2SCAN$ optimized structure at 700 GPa.

**Table S1.** Dependence of the calculated band gaps on the level of theory and pressure. Band gaps were predicted only for the *C2/c* phase. All the other phases were predicted to be metallic.

| Functional | Phase | Pressure [GPa] | Band-gap [eV] |
| --- | --- | --- | --- |
| $R^2SCAN$ | *C2/c* | 400 | 0.408 |
| SCAN0 | *C2/c* | 400 | 1.747 |
| SCAN0 | *C2/c* | 500 | 0.794[a] |

[a] A phase transition to the *Cmca-12* phase, followed by metallization is predicted at 470 GPa, in agreement with the experimental observation (~425 GPa). [32]

## S5. Phonon Band Structures

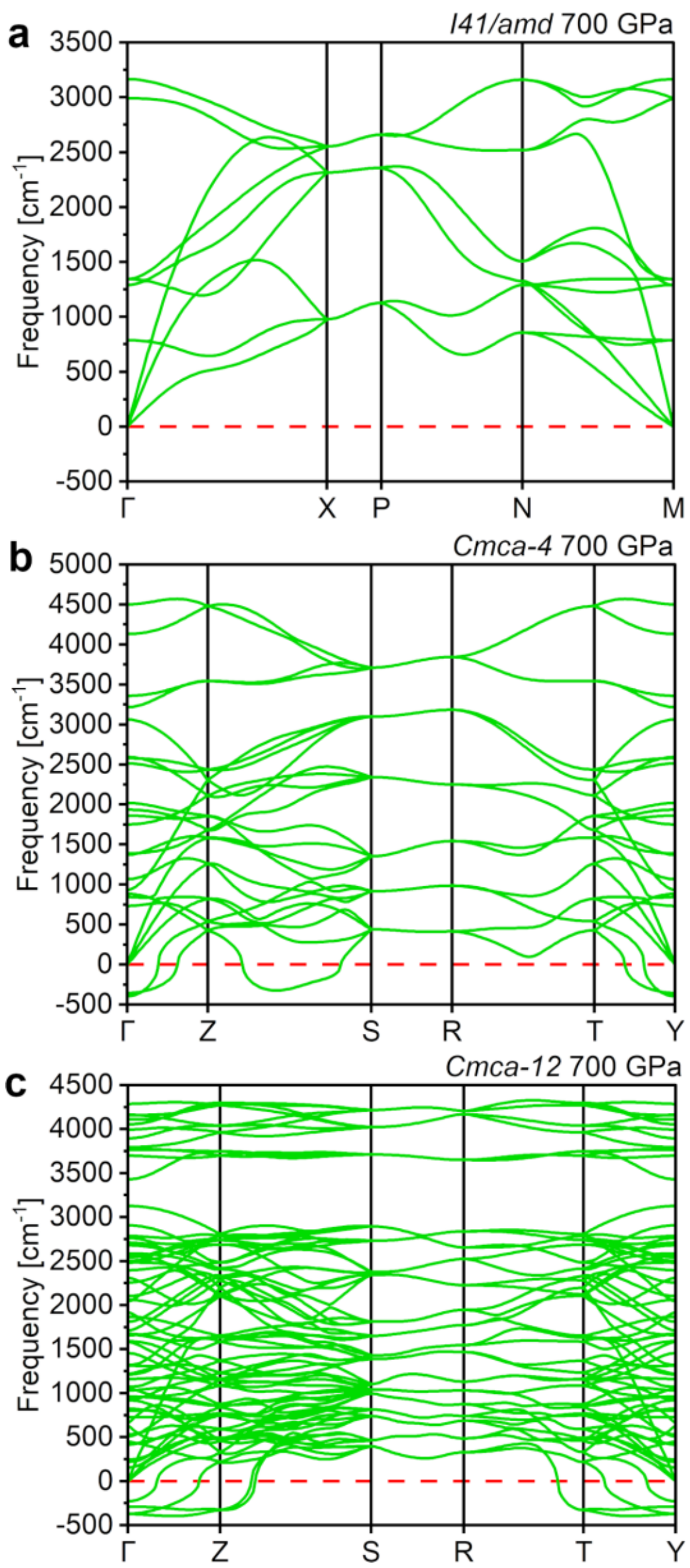

**Figure S8.** Phonon band structures of the optimized *$I4_1/amd$*, *Cmca-4*, *Cmca-12* and *C2/c* phases, calculated with the PBE functional at $P$ = 700 GPa with VASP (finite displacement method) on the geometries optimized with the $R^2$SCAN functional.